\documentclass[aps,pra,twocolumn,amsmath,nofootinbib]{revtex4-1}
\usepackage{epsfig}
\usepackage{graphics}
\usepackage{color}
\usepackage{epstopdf}
\usepackage{bm}
\usepackage{graphicx}
\usepackage{dcolumn}
\usepackage{cases}
\usepackage{appendix}
\usepackage{ulem}

\begin{document}
\author{Meng-Yun Lai$^{1}$}
\author{Yong-Long Wang$^{2,3,4}$}~\email[]{wangyonglong@lyu.edu.cn}
\author{Guo-Hua Liang$^{5}$}
\author{Fan Wang$^{1}$}
\author{Hong-Shi Zong$^{1,6,7}$}~\email[]{zonghs@nju.edu.cn}
\address{$^{1}$Department of Physics, Nanjing University, Nanjing 210093, China}
\address{$^{2}$National Laboratory of Solid State Microstructures, Department of Materials Science and Engineering, Nanjing University, Nanjing 210093, China}
\address{$^{3}$Collaborative Innovation Center of Advanced Microstructures, Nanjing University, Nanjing 210093, China}
\address{$^{4}$School of Physics and Electronic Engineering, Linyi University, Linyi, 276005, China}
\address{$^{5}$College of engineering and applied sciences, Nanjing University, Nanjing 210093, China}
\address{$^{6}$Joint Center for Particle, Nuclear Physics and Cosmology, Nanjing 210093, China}
\address{$^{7}$State Key Laboratory of Theoretical Physics, Institute of Theoretical Physics, CAS, Beijing, 100190, China}


\title{Electromagnetic wave propagating along a space curve}
\begin{abstract}
 Using the thin-layer approach, we derive the effective equation for the electromagnetic wave propagating along a space curve. We find intrinsic spin-orbit, extrinsic spin-orbit and extrinsic orbital angular momentum and intrinsic orbital angular momentum couplings induced by torsion, which can lead to geometric phase, spin and orbital Hall effects. And we show the helicity inversion induced by curvature that can convert the right-handed circularly polarized electromagnetic wave into left-handed polarized one, vice verse. Finally, we demonstrate that the gauge invariance of the effective dynamics is protected by the geometrically induced gauge potential.
\end{abstract}


\maketitle


\section{Introduction}

It is known that an electromagnetic wave can have three types of angular momentum (AM), they are spin angular momentum (SAM), intrinsic orbital angular momentum (IOAM) and extrinsic orbital angular momentum (EOAM), respectively \cite{franke2008advances,BLIOKH2015physrep,Bliokh2015Natpho}. SAM is determined by the polarization of light, IOAM is associated with the optical vortices, and EOAM relates to the trajectory of light \cite{Bliokh2015Natpho}. In recent years, the interactions among the three angular momenta (AMI) are attracting growing attention \cite{Onoda2004,Berry2005, Marrucci2006,zhao2007,Hosten787,Bliokh2008,rodriguez2010,Vuong2010,Lin331,Shitrit724, Petersen67,O’Connor2014,Bliokh1448}, due to both theoretical interest and potential applications in modern electromagnetics. For example, the intrinsic and extrinsic spin-orbit couplings lead to the geometrical phases and Hall effects \cite{Bliokh2006PRL}. These optical AMI phenomena manifest the inherent topological features of electromagnetic wave. Importantly, the topological features become conspicuous when electromagnetic wave propagates along a curved path \cite{Ross1984,Haldane:86,Chiao1986,berry1987interpreting,Bliokh2008,Bliokh2009JOA,Bliokh2009PRA}, in which the AMIs of electromagnetic wave can be manifested by the geometry of the curve.

In most of cases, since the traditional scalar approximation can not adequately describe the AMI phenomena of the polarized electromagnetic wave, a thorough vector analysis becomes necessary. For instance, by using the geometrical optics of vector waves, the Berry phase and spin and orbital Hall effects were explained in terms of the Coriolis effect \cite{Bliokh2009JOA,Bliokh2009PRA}. Another  suitable candidate to provide such analysis is the thin-layer approach. The thin-layer approach is an intuitive framework to study various types of waves that constrained to a curved surface or a curve \cite{Costa1981,burgess1993,Ouyang1999,Longhi:07,batz2008,TAIRA2011, yong-longwang2014,yong-longWang2017}. In particular, the thin-layer approach has been used to study the evolutions of electromagnetic wave constrained to a curved surface and a curve widely in the past decade \cite{Longhi:07,batz2008,Schultheiss2010,TAIRA2011,Bekenstein2017}. Experimentally, the curved surface and curve can be (say) a curved film waveguide and a curved optical fiber respectively. We know that light consists of electromagnetic waves. When an electromagnetic wave is constrained to a curved surface, a curvature-induced scalar potential arises \cite{batz2008}. It was theoretically and experimentally shown that the change of the curvature is equivalent to that of the refractive index \cite{Longhi:07,Schultheiss2010}. When an electromagnetic wave is constrained to a curve, the torsion of the curve plays the role of an effective gauge potential. In Ref.~\cite{TAIRA2011}, the scalar approximation for the electromagnetic field was adopted, it was shown that the torsion-induced gauge potential results in IOAM-dependent geometrical phase. Furthermore, for an electromagnetic wave constrained to a space curve the induced AMIs need to study in an effective vector formalism.

In order to study the optical AMIs by the geometry of a space curve, we provide a full-vector analysis of the effective equation for electromagnetic wave propagating along the curve in the thin-layer formalism. The present paper is organized as follows. In sec.\ref{EffectiveFieldEq}, we derive the effective equation describing the propagation of electromagnetic wave along a space curve. In sec.\ref{GeometryInducedEffects}, we analyze some geometry-induced effects. Sec.\ref{gauge} provides  the gauge analysis for the effective equation. Sec.\ref{conclusions} gives conclusions.
\section{Effective equation}\label{EffectiveFieldEq}
Describing the propagation of electromagnetic waves without sources, the Maxwell equations in a general coordinates system (CS) read \cite{batz2008,landau2013classical}
\begin{eqnarray}
 \label{MaxwellEq1} \eta^{ijk}\partial_j E_k +  \frac{1}{c}\frac{\partial{ B^i}}{\partial t}&=& 0, \\
 \label{MaxwellEq2} \eta^{ijk}\partial_j H_k -  \frac{1}{c}\frac{\partial{ D^i}}{\partial t}&=& 0, \\
 \label{MaxwellEq3} \frac{1}{\sqrt{g}}\partial_i(\sqrt{g}g^{ij}B_j) &=& 0, \\
 \label{MaxwellEq4} \frac{1}{\sqrt{g}}\partial_i(\sqrt{g}g^{ij}D_j) &=& 0,
\end{eqnarray}
where the Latin indices $i, j, k$ run from 1 to 3, $\eta_{ijk}=\sqrt{g}\epsilon_{ijk}$ denote the covariant coordinate components of the invariant volume element, $\eta^{ijk}=(1/\sqrt{g})\epsilon^{ijk}$ the contravariant components, $g_{ij}$ the coordinate components of the Euclidean metric tensor and $g=det(g_{ij})$ the determinant of the metric \cite{nakahara2003geometry}. Here $\epsilon_{ijk}$ are the coordinate components of the invariant volume element in the Cartesian CS, also known as the Levi-Civita symbol.

From Eqs.~\eqref{MaxwellEq1} and \eqref{MaxwellEq2}, we can obtain the equation for electric field
\begin{eqnarray}\label{d'alembertEq}
  (\nabla^2-\frac{n^2}{c^2}\partial_t^2) {\vec{\bf E}} &=& 0,\\\nonumber
\end{eqnarray}
where $\nabla_i$ denotes the covariant derivative operator. This is a d'Alembert-type equation for a vector field. According to the differential geometry \cite{nakahara2003geometry}, Eq.~\eqref{d'alembertEq} can be expanded as
\begin{eqnarray}\label{FieldEq0}
  \frac{1}{\sqrt{g}}\partial_j(\sqrt{g}g^{jk}\partial_kE^i)  + \frac{1}{\sqrt{g}}\partial_j(\sqrt{g}g^{jk}\Gamma^i_{kl}E^l)\nonumber\\
  + g^{jk}\Gamma^i_{jl}\partial_kE^l+ g^{jk}\Gamma^i_{jl}\Gamma^l_{km}E^m-\frac{n^2}{c^2}\partial_t^2 E^i&=&0, \nonumber\\
\end{eqnarray}
where $\Gamma^i_{jk}$ is a Christoffel symbol. In contrast to the equation for a scalar field~\cite{TAIRA2011}, here appears three newly additional terms, they represent the unique features of vector field.

\begin{figure}
  \centering
  \includegraphics[width=0.35\textwidth]{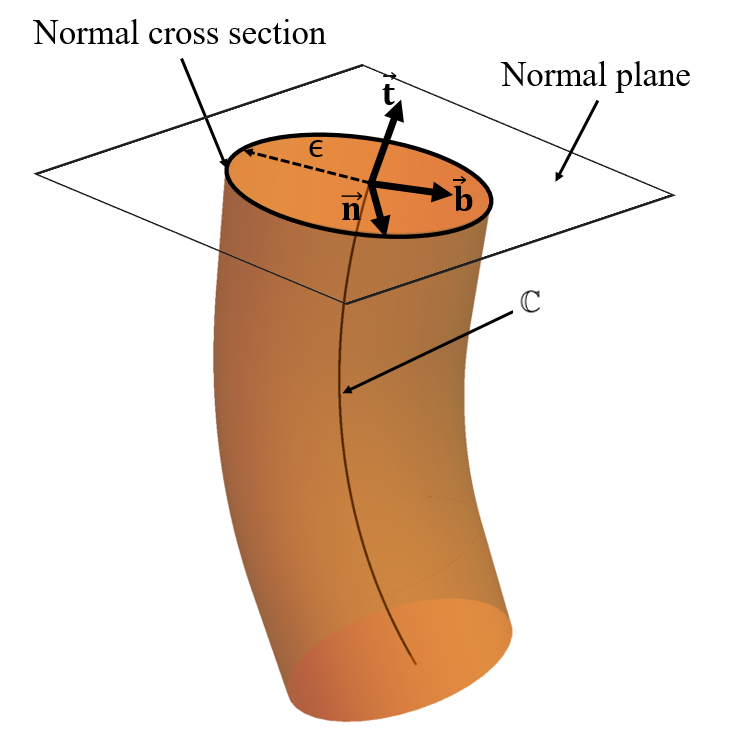}
  \caption{Schematic diagram of the thin tube. ${\vec{\bf t}}$, ${\vec{\bf n}}$ and ${\vec{\bf b}}$ are the tangent, normal and binormal vectors of the curve $\mathcal{C}$ respectively. $\epsilon$ is the radius of the normal cross section.}\label{001}
\end{figure}

Subsequently, we derive the effective equation on a curve from Eq.~\eqref{FieldEq0} in the thin-layer formalism. To do this, we need to introduce a CS describing the points in the neighborhood
of a curve $\mathcal{C}$ \cite{SCHUSTER2003132}. This CS is named as thin-tube CS in this paper. The neighborhood of $\mathcal{C}$ is denoted by $\mathcal{N}_c$, in which the points can be described by the following position vector
\begin{eqnarray}\label{positionvec1}
  {\vec{\bf R}}(s,q^2,q^3) &=& {\vec{\bf r}}(s) +q^2{\vec{\bf n}}(s) +q^3{\vec{\bf b}}(s),
\end{eqnarray}
where $\vec{\bf r}(s)$ parameterizes the curve $\mathcal{C}$, $s$ is the arc-length, $\vec{\bf n}$ and $\vec{\bf b}$ are the normal and binormal unit vectors of $\mathcal{C}$, $q^2$ and $q^3$ are the corresponding coordinate variables. The metric in the thin-tube CS is defined by
\begin{eqnarray}
  g_{ij} &=& \partial_i{\vec{\bf R}}\cdot\partial_j{\vec{\bf R}},
\end{eqnarray}
where the three coordinate variables are $\{q^1= s, q^2, q^3\}$ and the corresponding basis vectors are $\{\partial {\vec{\bf R}}/\partial q^i\}$. Besides the CS defined in $\mathcal{N}_c$, we also need a frame attached to $\mathcal{C}$. The well-known Frenet frame is a convenient choice. The Frenet frame is spanned by the three unit vectors attached to $\mathcal{C}$, $\{{\vec{\bf t}}(s)=\partial {\vec{\bf r}}/\partial s, {\vec{\bf n}}(s), {\vec{\bf b}}(s)\}$, where ${\vec{\bf t}}$ is the tangent vector of $\mathcal{C}$. According to the Frenet-Serret formulas, these vectors and their derivatives obey the following equation \cite{wang2017quantum},
\begin{eqnarray}\label{F-Sformulas}
  \left(
     \begin{array}{c}
       \partial_s {\vec{\bf t}}\\
       \partial_s {\vec{\bf n}} \\
       \partial_s {\vec{\bf b}} \\
     \end{array}
   \right)
   &=& \left(
          \begin{array}{ccc}
            0 & \kappa(s) & 0 \\
            -\kappa(s) & 0 & \tau(s) \\
            0 & -\tau(s) & 0 \\
          \end{array}
        \right)
        \left(
                 \begin{array}{c}
                   {\vec{\bf t}} \\
                   {\vec{\bf n}} \\
                   {\vec{\bf b}} \\
                 \end{array}
               \right),
\end{eqnarray}
where $\kappa(s)$ and $\tau(s)$ are the curvature and torsion of $\mathcal{C}$ respectively. The curvature and torsion describe the embedding of $\mathcal{C}$ in the three-dimensional Euclidean space. Then, the metric $g_{ij}$ can be expressed in terms of the curvature and torsion as
\begin{eqnarray}\label{metric}
  g_{ij} &=& \frac{\partial {\vec{\bf R}}}{\partial q^i}\cdot\frac{\partial {\vec{\bf R}}}{\partial q^j} \nonumber\\
  &=& \left(
  \begin{array}{ccc}
   g_{11} & -\tau(s)q^3 & \tau(s)q^2 \\
  -\tau(s)q^3 & 1 & 0 \\
  \tau(s)q^2 & 0 & 1 \\
  \end{array}
  \right),
\end{eqnarray}
where $g_{11}=[1-\kappa(s)q^2]^2+\tau(s)^2[(q^2)^2+(q^3)^2]$. Note that the term $1-\kappa(s)q^2$ in $g_{11}$ shows the asymmetry between ${\vec{\bf n}}$ and ${\vec{\bf b}}$, because of the bending of $\mathcal{C}$ along a specific direction in the normal plane. This asymmetry would lead to different evolutions for $E^2$ and $E^3$.

For the sake of simplicity, we consider a monochromatic transverse electric field ${\vec{\bf E}}=\{0,E^2,E^3\}$, with frequency $\omega=k_0c$, that is restricted in a curved thin tube (see Fig.~\ref{001}). The refractive index in the thin tube is nearly a constant \cite{berry1987interpreting,TAIRA2011}. In terms of the fundamental framework of the thin-layer formalism discussed in \cite{SCHUSTER2003132} and \cite{wang2017quantum}, we conclude the main steps of deriving the effective equation for electromagnetic wave propagating along $\mathcal{C}$ from Eq.~\eqref{FieldEq0}. First, to separate the tangent part of the electric field~\cite{Costa1981} by introducing a new wave $\bar{E}^i=g^{1/4}E^i$ \cite{batz2008}, where $g^{1/4}$ is a rescale factor that is determined by the right volume measure on $\mathcal{C}$. The usual scalar geometrical potential $\kappa^2/4$ results from the action of normal derivatives on $g^{-1/4}$ \cite{wang2017quantum}. Second, by integrating over the normal cross-section of the thin tube and transforming from linearly polarized basis to circularly polarized basis, we project Eq.~\eqref{FieldEq0} onto the direct product space $\mathcal{S}_{s}\otimes\mathcal{S}_o$, where $\mathcal{S}_{s}$ and $\mathcal{S}_o$ denote the spin space and a degenerate subspace of IOAM (we will discuss the two spaces below). After the projecting, the spin and orbital momentum operators becomes c-numbers. This makes the decouple between the normal and tangent parts of the wave equation possible \cite{SCHUSTER2003132}. At last, we impose the limiting $\epsilon\rightarrow 0$ to obtain the final effective equation, where $\epsilon$ is the radius
of the normal cross-section of the thin tube. In practice, the limiting can be substituted by the thin-tube condition
\begin{eqnarray}\label{ThinLayerCondition}
  \kappa^{-1},\tau^{-1} \gg \epsilon.
\end{eqnarray}
Rigorously, the derivatives of the curvature and torsion should be taken into account in the thin-tube condition. However, we do not study the effects caused by these derivatives. We suppose that the curvature and torsion of $\mathcal{C}$ are slowly varying.

Substituting $\bar{E}^i=g^{1/4}E^i$ and the metric Eq.~\eqref{metric} into Eq.~\eqref{FieldEq0}, we obtain
\begin{widetext}
\begin{eqnarray}\label{FieldEq2}
  g^{-1/2}\bigg[\partial_2(g^{1/2}\partial_2)+\partial_3(g^{1/2}\partial_3)+
\big[[\partial_s-
   i\tau(s)(\hat{L}_s+\hat{\Sigma}_s)]g^{-1/2}[\partial_s-
   i\tau(s)(\hat{L}_s+\hat{\Sigma}_s)]\big]\bigg]\left(
    \begin{array}{c}
      g^{-1/4}\bar{E}^2\\
      g^{-1/4}\bar{E}^3 \\
    \end{array}
  \right)\nonumber\\
  -g^{-2}\left(
    \begin{array}{cc}
      \kappa(s)^2 & 0 \\
      0 & 0 \\
    \end{array}
  \right)
  \left(
    \begin{array}{c}
      g^{-1/4}\bar{E}^2 \\
      g^{-1/4}\bar{E}^3 \\
    \end{array}
  \right)+k^2\left(
    \begin{array}{c}
      g^{-1/4}\bar{E}^2 \\
      g^{-1/4}\bar{E}^3 \\
    \end{array}
  \right)+O(\delta) \left(
    \begin{array}{c}
      g^{-1/4}\bar{E}^2 \\
      g^{-1/4}\bar{E}^3 \\
    \end{array}
  \right)=0,
\end{eqnarray}
\end{widetext}
where $\hat{L}_s=-i(q^2\partial_3-q^3\partial_2)=-i\partial_{\phi}$, $\hat{\Sigma}_s=g^{-1/2}\hat{\sigma}_s$, $\delta$ denotes $\max(\rho/\kappa^{-1}, \rho/\tau^{-1})$, $k=k_0n$ is the wave vector, the polar coordinates $\{\rho,\phi\}$ are defined by $\{q^2=\rho\cos\phi, q^3=\rho\sin\phi\}$ and $\hat{\sigma}_s$ is part of the spin-1 matrix operator $\hat{S}_3$ with
\begin{eqnarray}
   \hat{\sigma}_s=\left(
    \begin{array}{cc}
      0 & -i \\
      i & 0 \\
    \end{array}
  \right), ~~~~~~\hat{S}_3=\left(
    \begin{array}{ccc}
      0 & -i & 0\\
      i & 0 & 0\\
       0 & 0 & 0\\
    \end{array}
  \right).
  \end{eqnarray}
In Eq.~\eqref{FieldEq2}, $\partial_i$ and $\hat{L}_s$ act on the spatial distribution of the electric field, whereas the $2\times2$ matrices act on $\left(
    \begin{array}{c}
      g^{-1/4}\bar{E}^2 \\
      g^{-1/4}\bar{E}^3 \\
    \end{array}
  \right)$.

With the thin-tube condition Eq.~\eqref{ThinLayerCondition}, we can use the ratio
$\delta$ as a natural perturbative parameter. Thus the dominant normal differential equation describing the normal electric field is
\cite{Białynicki1994,SCHUSTER2003132}
\begin{eqnarray}\label{NormalEq1}
[\frac{1}{\rho}\frac{\partial}{\rho}(\rho\frac{\partial}{\rho})+
\frac{1}{\rho^2}\frac{\partial^2}{\partial\phi^2}-k_{\bot}^2]\bar{E}_{\bot}(\rho,\phi)=0,
\end{eqnarray}
where $k_{\bot}^2=k^2-k_s^2$, $k_{\bot}$ and $k_s$ are normal and tangent wave vectors, respectively.
In Eq.~\eqref{NormalEq1} the separation of the electric field into tangent and normal components is considered, $\bar{E}(s,\rho,\phi)=\bar{E}_s(s)\bar{E}_{\bot}(\rho,\phi)$.
With a certain value of $l$, Eq.~\eqref{NormalEq1} can give exact two-fold degenerate solutions
\begin{eqnarray}\label{DegenerateSolutions}
\bigg\{\begin{array}{cc}
\bar{E}_{\perp}(\rho,\phi;|l|) \propto e^{i|l|\phi}, \\\\
\bar{E}_{\perp}(\rho,\phi;-|l|) \propto e^{-i|l|\phi},
          \end{array}
\end{eqnarray}
where $l=0,\pm1,\pm2,\pm3...$ are the eigenvalues for $\hat{L}_s$. The solutions Eq.~\eqref{DegenerateSolutions} span a two-dimensional degenerate subspace $\mathcal{S}_o$. The direct product of the degenerate subspace  $\mathcal{S}_o$ and the spin space $\mathcal{S}_{s}$ gives a four-dimensional space. The basis for the four-dimensional space is
 \begin{eqnarray}\label{BasisFor4}
  \left(
  \begin{array}{c}
     {\vec{\bf e}}_+\bar{E}_{\perp}(\rho,\phi;|l|) \\
    {\vec{\bf e}}_+\bar{E}_{\perp}(\rho,\phi;-|l|) \\
     {\vec{\bf e}}_-\bar{E}_{\perp}(\rho,\phi;|l|) \\
    {\vec{\bf e}}_-\bar{E}_{\perp}(\rho,\phi;-|l|) \\
  \end{array}
\right),
 \end{eqnarray}
 where the basis for spin space are defined by
\begin{eqnarray}
  {\vec{\bf e}}_+ &=& \frac{1}{\sqrt{2}}({\vec{\bf n}}+i{\vec{\bf b}}), \nonumber\\
   {\vec{\bf e}}_- &=& \frac{1}{\sqrt{2}}({\vec{\bf n}}-i{\vec{\bf b}}).
\end{eqnarray}
 By substituting $\bar{E}^{i}(s,\rho,\phi;l)=\bar{E}^{(i,l)}_s(s)\bar{E}_{\perp}(\rho,\phi;l)$
 into Eq.~\eqref{FieldEq2} and applying the integration
 $\int^\epsilon_0\int^{2\pi}_0  \rho {\rm d}\rho{\rm d} \phi ~\bar{E}^{\ast}_{\perp}(\rho,\phi;l)$, Eq.~\eqref{FieldEq2} can be projected onto the degenerate subspace $\mathcal{S}_o$.
 Then by performing the following transformation
\begin{eqnarray}\label{FieldEqOnC}
  \left(
    \begin{array}{c}
      \bar{E}^+ \\
      \bar{E}^- \\
    \end{array}
  \right)
   &=&\frac{1}{\sqrt{2}}\left(
        \begin{array}{cc}
          1 & -i \\
          1 & i \\
        \end{array}
      \right)
    \left(
         \begin{array}{c}
      \bar{E}^2 \\
      \bar{E}^3 \\
         \end{array}
       \right),
\end{eqnarray}
Eq.~\eqref{FieldEq2} can be projected onto the spin space $\mathcal{S}_s$. Finally, using the thin-tube condition, we obtain the effective equation on the curve
\begin{widetext}
\begin{eqnarray}\label{FieldEqOnC0}
\bigg[\big[\partial_s -i(l+\hat{\sigma})\tau\big]^2
-\frac{1}{4}\kappa^2
-\frac{1}{2}\kappa^2\left(
 \begin{array}{cc}
 0 & 1 \\
 1 & 0 \\
 \end{array}
 \right)
+k^2_s\bigg]\left(
\begin{array}{cc}
\bar{E}^{(+,l)}_s \\ \bar{E}^{(-,l)}_s \\
\end{array}
\right)=0,
\end{eqnarray}
\end{widetext}
where $\hat{\sigma}=diag(1,-1)$ denotes the helicity operator of electromagnetic wave.  Eq.~\eqref{FieldEqOnC0} can be rewritten in the following compact form
\begin{eqnarray}\label{FieldEqOnC1}
\big[[\partial_s -i(l+{\sigma})\tau]^2
-\frac{1}{4}\kappa^2
+k^2_s\big]
\bar{E}^{(\sigma,l)}_s -\frac{1}{2}\kappa^2\bar{E}^{(-\sigma,l)}_s =0,\nonumber\\
\end{eqnarray}
where ${\sigma}=\pm1$ denotes the helicity of electromagnetic wave.
These two equivalent equations are the key results of the present paper and all the following analyses are based on them.
It is worth pointing out that the effective Hamiltonian evolution equation in Ref.~\cite{berry1987interpreting} can be derived from
Eq.~\eqref{FieldEqOnC1}.

Interestingly, there are three additional terms induced by torsion and curvature in Eq.~\eqref{FieldEqOnC1}. The first term, $-i(l+\sigma)\tau$ is equivalent to an effective gauge term, which will be discussed in Sec.~\ref{gauge}. The torsion $\tau$ plays the role of an effective gauge potential. As $\tau$ is a function of $s$, it is equivalent to a local gauge potential. For a constant, it becomes global. Analogous to the
Aharonov-Bohm effect, $\tau$ results in geometrical phases. Furthermore, $\tau$ provides a way to control the spin and the vortex of electromagnetic wave due to the SAM-EOAM and IOAM-EOAM couplings $-2\tau(l+\sigma)i\partial_s$, and the SAM-IOAM coupling $-2\tau^2l\sigma$. The second term $-\kappa^2/4$ is a scalar potential that can be taken as a correction to the index $n^2$. The last term is induced by the curvature and multiplied by an off-diagonal matrix. As mentioned above, it results from the asymmetry between ${\vec{\bf n}}$ and ${\vec{\bf b}}$ caused by the bending of $\mathcal{C}$. Obviously,
the off-diagonal matrix results in the helicity inversion.

\section{geometry-induced effects}\label{GeometryInducedEffects}
In general, the wavelength of electromagnetic wave is much smaller than the size of the curved thin-tube, i.e.,
\begin{eqnarray}\label{approximateCondition}
 k_s &\gg& \tau,\kappa.
\end{eqnarray}
Therefore, we can define a dimensionless parameter as
\begin{eqnarray}
  \Delta &=& \max(\frac{\lambda_s}{R_{\kappa}}, \frac{\lambda_s}{R_{\tau}}),
\end{eqnarray}
where $\lambda_s=2\pi/{k_s}$, $R_{\kappa}=\kappa^{-1}$ and $R_{\tau}=\tau^{-1}$. As mentioned above, the gauge term in Eq.~\eqref{FieldEqOnC1} induces the couplings between different angular momenta. The couplings can lead to various interesting phenomena. Especially, the SAM-EOAM and IOAM-EOAM couplings result in two mutual phenomena: the geometrical phase and spin and orbital Hall effects. The magnitudes of the two phenomena are of the orders of $\Delta^{0}$ and $\Delta$ respectively. The geometrical phase are represented by
\begin{eqnarray}\label{SolutionwaveEqs4}
 \bar{E}^{(\sigma,l)}_{s} &\propto&  \exp\big[i(l+\sigma)\gamma(s)\big],
\end{eqnarray}
where $\gamma(s)=\int^s_0\tau(s')ds'$. Eq.~\eqref{SolutionwaveEqs4} indicates that the electromagnetic wave acquires a geometrical phase factor $\exp\big[i(l+\sigma)\gamma(s)\big]$ after the propagation along a curve. This geometrical phase consists of two components, SAM-dependent and IOAM-dependent. The SAM-dependent one coincides with the well-known Berry phase in optics \cite{Chiao1986}. On the other hand, the spin and orbital Hall effects represent the SAM-dependent and IOAM-dependent corrections to the ray trajectory of electromagnetic wave \cite{Bliokh2006PRL,Bliokh2008}. Similar to the spin Hall effect of light \cite{Bliokh2009JOA}, the SAM-dependent and IOAM-dependent corrections are given by \begin{eqnarray}\label{deltar1}
  \delta {\vec{\bf r}} &=& -\int\lambdabar(l+\sigma)\kappa{\rm d}s~{\vec{\bf b}},
 \end{eqnarray}
where $\lambdabar=1/k$. While the geometrical phase represents the influences of the curve on the SAM and IOAM of electromagnetic wave, the Hall effects show the effects of SAM and IOAM of electromagnetic wave on the ray trajectory.

\begin{figure}
  \centering
  \includegraphics[width=0.4\textwidth]{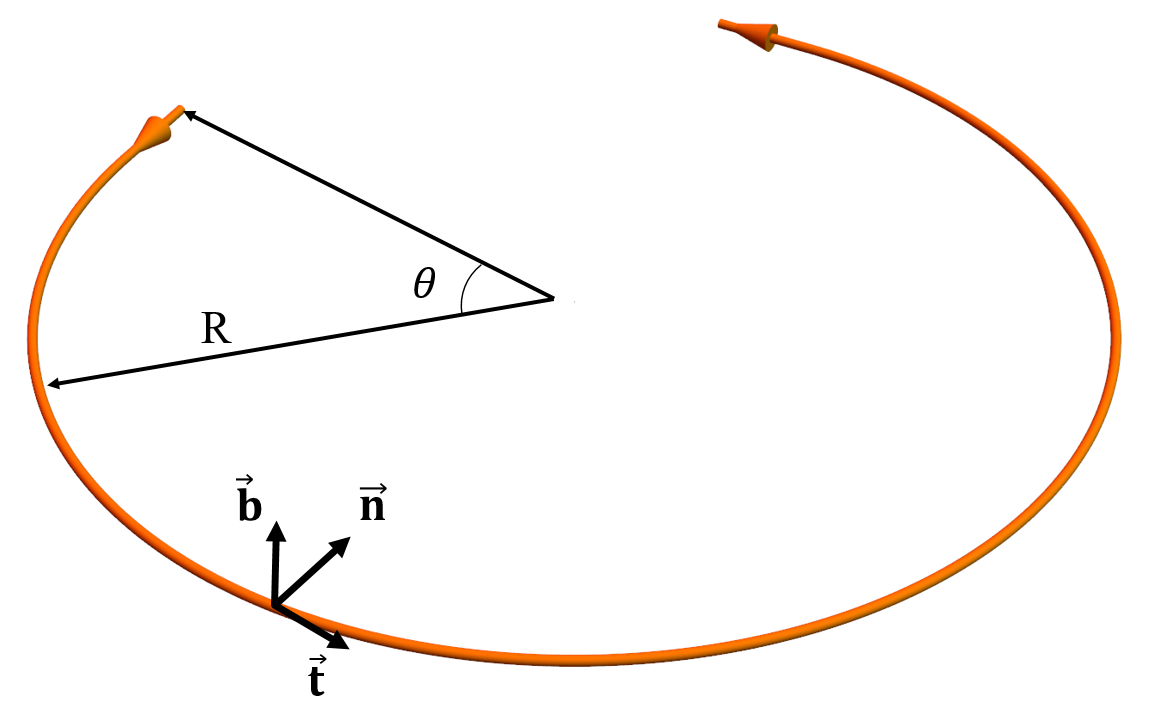}
  \caption{A planar curve with constant radius $R$. ${\vec{\bf t}}$, ${\vec{\bf n}}$ and ${\vec{\bf b}}$ are the tangent, normal and binormal vectors of this curve respectively.}\label{002}
\end{figure}

Once the typical sizes of the curve are comparable to the wavelength scale, the $\Delta^2$-order effect would become important.
In Eq.~\eqref{FieldEqOnC1}, the $\Delta^2$-order effect is described by $\kappa^2$. To highlight the
curvature-induced effect, we consider a torsion-free curve (see Fig.~\ref{002}). From Eq.~\eqref{FieldEqOnC1}, in terms of the linearly polarized basis,
 the field equations for $\bar{E}^n_s$ ($\bar{E}^2_s$) and $\bar{E}^b_s$ ($\bar{E}^3_s$) become
\begin{eqnarray}\label{example1}
  \partial_s ^2\bar{E}_s^n +K_n^2\bar{E}^n_s &=& 0, \\\label{example2}
  \partial_s ^2\bar{E}_s^b +K_b^2\bar{E}^b_s &=& 0,
\end{eqnarray}
where $K_n=\sqrt{k^2_s -3\kappa^2/4}$ and $K_b=\sqrt{k^2_s +\kappa^2/4}$. The difference between $K_n$ and $K_b$ shows that $\bar{E}^n_s$ and $\bar{E}^b_s$ have different propagations.

For simplicity, we further assume that the radius $R$ of the planar curve is a constant, then we have the total arc-length $L=R\theta$, the curvature $\kappa=1/R$ and $s\in[0,L]$ (see Fig.~\ref{002}). Therefore, analogous to the one dimensional scattering problem
in quantum mechanics, the transmission coefficients for
$\bar{E}^n_s$ and $\bar{E}^b_s$ can be easily obtained
\begin{eqnarray}\label{transmissioncoefficients}
  T_n &=& \bigg\{\begin{array}{cc}
            \frac{1}{1+\frac{1}{4}(\frac{k_s}{K_n}-\frac{K_n}{k_s})^2\sin^2(K_nL)},
            & ~~~k^2_s > \frac{3\kappa^2}{4}, \\
            \frac{1}{1+\frac{1}{4}\frac{(k^2_s+K_n'^2)^2}{k^2_sK_n'^2} {\rm sh}
            ^2(K_n'L)}, & ~~~k^2_s < \frac{3\kappa^2}{4},
          \end{array}
\\
  T_b &=& \frac{1}{1+\frac{1}{4}(\frac{k_s}{K_b}-\frac{K_b}{k_s})^2\sin^2(K_bL)},
\end{eqnarray}
where $K'_n=iK_n$, $T_n$ and $T_b$ are the transmission coefficients for
$\bar{E}^n_s$ and $\bar{E}^b_s$ respectively.
Even for $R\sim10\lambda_s$ $1-T_n$ and $1-T_b$ never exceed $10^{-8}$. The solutions for Eqs.~\eqref{example1}
and \eqref{example2} are
\begin{eqnarray}\label{solutionsforen}
  \bar{E}^n_s &=& \bigg\{\begin{array}{cc}
            A_ne^{iK_ns}+B_ne^{-iK_ns}, & ~~~k^2_s > \frac{3\kappa^2}{4}, \\
            A_n'e^{K_n's}+B_n'e^{-K_n's}, & ~~~k^2_s < \frac{3\kappa^2}{4},
          \end{array} \\\label{solutionsforeb}
  \bar{E}^b_s  &=& A_be^{iK_bs}+B_be^{-iK_bs},
\end{eqnarray}
with
\begin{eqnarray}\label{solutionsforeneb}
  A_n &=& \frac{\sqrt{T_n}}{2}(1+\frac{k_s}{K_n})e^{i(k_s-K_n)L}, \nonumber\\ B_n& =&
  \frac{\sqrt{T_n}}{2}(1-\frac{k_s}{K_n})e^{i(k_s+K_n)L}, \nonumber\\
  A_n' &=& \frac{\sqrt{T_n}}{2}(1+\frac{ik_s}{K_n'})e^{i(k_s+iK_n')L}, \nonumber\\B_n'
  &= & \frac{\sqrt{T_n}}{2}(1-\frac{ik_s}{K_n'})e^{i(k_s-iK_n')L}, \nonumber\\
  A_b &=& \frac{\sqrt{T_b}}{2}(1+\frac{k_s}{K_b})e^{i(k_s-K_b)L}, \nonumber\\B_b &= &
  \frac{\sqrt{T_b}}{2}(1-\frac{k_s}{K_b})e^{i(k_s+K_b)L}.
\end{eqnarray}
Even for $R\sim10\lambda_s$ the coefficients of the reflected waves, $B_n$ and $B_b$, never exceed $10^{-4}$. Therefore, the reflected waves can be safely neglected. In other words, $\kappa^2$ can be neglected, that is $K_n\approx K_b\approx k_s$ being good approximations. Thus Eqs.~\eqref{example1} and \eqref{example2} reduce to Helmholtz equations. The helicity of electromagnetic wave remains adiabatic invariant along the curve. When $\kappa^2$ can not be neglected but is still smaller than $k^2_s$, the solutions Eqs.~\eqref{solutionsforen}, \eqref{solutionsforeb} and \eqref{solutionsforeneb} give the probability for a change from $\sigma=+$ to
  $\sigma=-$
\begin{eqnarray}\label{approximateeq}
  P_{+-} &=& \sin^2(\frac{\kappa^2s}{4k_s}).
\end{eqnarray}
This result is in nice agreement with that given in Ref.~\cite{berry1987interpreting}. Eq.~\eqref{approximateeq} indicates that the period of the helicity inversion of electromagnetic wave is $8\pi^2R^2/\lambda_s$. In particular, when the total length of the curve is $L=4\pi^2R^2(2m+1)/\lambda_s$, $m=0,1,2...$, the propagation along the curve can entirely converts a right-handed circularly polarized electromagnetic wave into a left-handed circularly polarized one. As an example, we consider the propagation through $N$ circles, the radius of each circle equals to $R$, then Eq.~\eqref{approximateeq} becomes
\begin{eqnarray}
  P_{+-} &=& \sin^2(\frac{\lambda_sN}{4R}).
\end{eqnarray}
For $R=10\lambda_s$, after the propagation through $N=10$ circles, about $6.12\%$ of the right-handed circularly polarized electromagnetic wave becomes left-handed circularly polarized.

\section{Gauge analysis}\label{gauge}
All the above calculations are performed in a special thin-tube CS, in which the normal basis vectors are defined in the Frenet frame. In what follows, we thus name it as Frenet thin-tube CS. In this CS, the basis vectors are  normal and binormal to $\mathcal{C}$. Note that an adapted frame can be obtained by rotating the Frenet frame~\cite{SCHUSTER2003132}. Therefore, the choice of normal basis vectors can not change the effective equation on the curve. In other words, the form of the effective equation depends on the symmetry of the normal part of the electric field. This can be displayed by the confinement boundary \cite{wang2017quantum}.
In the present paper, since we have assumed that the refractive index is constant and the normal boundary is circular, the gauge invariance of the effective equation is preserved after the thin-layer procedure.

To better understand the gauge structure of the effective equation on $\mathcal{C}$, we need to work in a general thin-tube CS. In fact, the general thin-tube CS can be defined by an arbitrary smooth assignment of the two basis vectors of the normal plane, i.e., the position vector becomes
\begin{eqnarray}\label{positionvec2}
  {\vec{\bf R}}(s,q^2,q^3) &=& {\vec{\bf r}}(s) +q^2{\vec{\bf v}}(s) +q^3{\vec{\bf w}}(s),
\end{eqnarray}
where $\{{\vec{\bf v}}, {\vec{\bf w}}\}$ are two orthonormal basis vectors of the normal plane. By applying the thin-layer procedure to Eq.~\eqref{FieldEq0}, we obtain the effective equation in the general thin-tube CS, that is
\begin{widetext}
\begin{eqnarray}\label{GeneralFieldEqOnC}
\bigg[\big[\partial_s -i(l+\sigma)\omega_t\big]^2
-\frac{1}{4}(\omega_v^2+\omega_w^2)
+k^2_s\bigg]
\bar{E}^{(\sigma,l)}_s -\frac{1}{2}(\omega_w-i\sigma\omega_v)^2\bar{E}^{(-\sigma,l)}_s=0,
\end{eqnarray}
\end{widetext}
where the three angular velocities $\omega_t$, $\omega_v$ and $\omega_w$ are defined by
\begin{eqnarray}
  \omega_t = {\vec{\bf w}}\cdot\partial_s{\vec{\bf v}}~~~~~~\omega_v = {\vec{\bf w}}\cdot\partial_s{\vec{\bf t}}
  ~~~~~~\omega_w = {\vec{\bf v}}\cdot\partial_s{\vec{\bf t}}.
\end{eqnarray}
These three angular velocities describe the rotations of the ${\vec{\bf v}}$-${\vec{\bf w}}$,
${\vec{\bf w}}$-${\vec{\bf t}}$ and ${\vec{\bf t}}$-${\vec{\bf v}}$ planes respectively, and are the generalizations of the curvature and torsion.
Consider a local rotation of the normal plane
\begin{eqnarray}\label{GaugeTransformation}
  \left(
    \begin{array}{c}
      \vec{\bf v} \\
      \vec{\bf w} \\
    \end{array}
  \right)\rightarrow  e^{i\theta(s)\hat{\sigma}_s}\left(
    \begin{array}{c}
      \vec{\bf v} \\
      \vec{\bf w} \\
    \end{array}
  \right).
\end{eqnarray}
Under the rotation, the three angular velocities transform as
\begin{eqnarray}\label{TransformationsForAngularvelocities}
  \omega_t &\rightarrow& \omega_t+\partial_s\theta,  \nonumber\\
  \omega_v &\rightarrow& -\sin\theta\omega_w+\cos\theta\omega_v, \nonumber\\
  \omega_w &\rightarrow& \cos\theta\omega_w+\sin\theta\omega_v,
\end{eqnarray}
and the corresponding transformation for the field $\bar{E}^{(\sigma,l)}_s$ is
\begin{eqnarray}\label{TransformationForField}
  \bar{E}^{(\sigma,l)}_s &\rightarrow& e^{i(l+\sigma)\theta(s)}\bar{E}^{(\sigma,l)}_s.
\end{eqnarray}
Then one can easily verify the gauge invariance of Eq.~\eqref{GeneralFieldEqOnC} by substituting Eqs.~\eqref{TransformationsForAngularvelocities} and ~\eqref{TransformationForField} into Eq.~\eqref{GeneralFieldEqOnC}.

It is worthwhile to point out that the gauge theory can be used to study the AMI of electromagnetic wave. For an electromagnetic wave propagating along a space curve, its wavevector is always in the tangent direction. The IOAM and SAM of electromagnetic wave are both defined in the plane normal to the curve. By limiting the scale size of the normal plane, the IOAM and SAM are mapped to the effective tangent dynamics. As we previously discussed, the local $SO(2)$ rotation of the normal plane should not changes the tangent dynamics. In this manner, each point on the curve associates with an independent internal $SO(2)$ group. For the electromagnetic wave with definite $l$ and $\sigma$, the electric field can take the following form
\begin{eqnarray}
 {\vec{\bf E}}(s,\rho,\phi;l,\sigma)={E}(s,\rho;l,\sigma)e^{il\phi}{\vec{\bf e}}_{\sigma}.
\end{eqnarray}
Under the local $SO(2)$ rotation Eq.~\eqref{GaugeTransformation}, ${E}(s,\rho;l,\sigma)$ is transformed as
\begin{eqnarray}\label{generalgauge2}
{E}(s,\rho;l,\sigma) &\rightarrow& e^{i(l+\sigma)\theta(s)}{E}(s,\rho;l,\sigma).
\end{eqnarray}
Subsequently, $\partial_s{E}(s,\rho;l,\sigma)$ is replaced by $[\partial_s-i(l+\sigma)\omega_t]\partial_s{E}(s,\rho;l,\sigma)$. The angular momenta $l+\sigma$ and the connection $\omega_t$ play the roles of the coupling constant and gauge potential, respectively. As a result, the couplings between different angular momenta arise. Suppose the general form for $\omega_v$ and $\omega_w$ in the tangent equation is
\begin{eqnarray}
  \left(
    \begin{array}{cc}
      f(\omega_v,\omega_w) & F(\omega_v,\omega_w) \\
      G(\omega_v,\omega_w) & g(\omega_v,\omega_w) \\
    \end{array}
  \right)\left(
\begin{array}{cc}
{E}(s,\rho;l,+) \\ {E}(s,\rho;l,-) \\
\end{array}
\right).
\end{eqnarray}
In light of the gauge invariance, the four functions should be transformed as
\begin{eqnarray}
  f(\omega_v,\omega_w) &\rightarrow& f(\omega_v,\omega_w),  \nonumber\\
  g(\omega_v,\omega_w) &\rightarrow& g(\omega_v,\omega_w),  \nonumber\\
  F(\omega_v,\omega_w) &\rightarrow& e^{2i\theta(s)}F(\omega_v,\omega_w),                         \nonumber\\
  G(\omega_v,\omega_w) &\rightarrow& e^{-2i\theta(s)}G(\omega_v,\omega_w).
\end{eqnarray}
Therefore, the simplest form of $f(\omega_v,\omega_w)$ and $g(\omega_v,\omega_w)$ is possibly the
"length" $(\omega_v^2+\omega_w^2)$. Under the condition, $F(\omega_v,\omega_w)$ and $G(\omega_v,\omega_w)$ must be
\begin{eqnarray}
  F(\omega_v,\omega_w) &\propto& (\omega_w-i\omega_v)^2, \\
  G(\omega_v,\omega_w) &\propto& (\omega_w+i\omega_v)^2.
\end{eqnarray}

As a consequence, the arbitrary assignment of the normal basis vectors implies a connection added to the gauge potential. Among the various adapted frames, the Frenet frame is a special one, in which the angular velocities (i.e., the torsion and curvature) describe the geometrical properties of the curve. Specifically, the torsion of the curve can not be eliminated via a local gauge transformation. In other words, the torsion provides a platform to show the gauge potential, which is measurable, physical. Therefore, an arbitrary gauge potential $\omega_t$ can be decomposed into two parts
\begin{eqnarray}
  \omega_t &=& \tau + \omega_{pure},
\end{eqnarray}
where $\tau$ stands for the physical component, $\omega_{pure}$ denotes the pure-gauge potential \cite{chen2008,Wang2015}.



\section{Conclusions}\label{conclusions}
We first employed the thin-layer approach to study the electromagnetic wave constrained to a curve and obtained the effective equation Eq.~\eqref{FieldEqOnC1}. In contrast with the Schr\"{o}dinger equation and the electromagnetic wave in scalar approximation \cite{wang2017quantum,TAIRA2011}, Eq.~\eqref{FieldEqOnC1} contains both the IOAM-related and SAM-related terms, which result from the vector nature of the electromagnetic wave. These terms are induced by curvature and torsion, and they can lead to the geometrical phase, spin and orbital Hall effects, and the helicity-dependent vortices for the electromagnetic wave constrained to propagate along a space curve~\cite{Bliokh2015Natpho}. Moreover, we found that the curvature-induced off-diagonal term can bring about the different evolutions for the electric fields that along normal and binormal directions, which results in the helicity inversion.

In addition, we examined the gauge invariance of the effective field equation, and demonstrated that the effective gauge term and off-diagonal term in the effective dynamics are inherent features, which are universal for electromagnetic wave. When electromagnetic wave propagates along a curve, the geometry of the curve provides a platform to manifest the gauge structure of the electromagnetic wave.

\acknowledgments
This work is supported in part by the National Natural Science Foundation of China (under Grants No. 11475085, No. 11535005, No. 11690030, No. 11625418 and No. 51721001), the Fundamental Research Funds for the Central Universities (under Grant No. 020414380074). Y.-L. W. was funded by the China Postdoctoral Science
Foundation (Grant No. 2017M611770), the Natural Science Foundation of Shandong Province of China (Grant No. ZR2017MA010) and Linyi University (LYDX2016BS135)

\normalem
\bibliographystyle{apsrev4-1}
\bibliography{QuantumEffectsOfCurvature}

\begin{thebibliography}{40}%
\makeatletter
\providecommand \@ifxundefined [1]{%
 \@ifx{#1\undefined}
}%
\providecommand \@ifnum [1]{%
 \ifnum #1\expandafter \@firstoftwo
 \else \expandafter \@secondoftwo
 \fi
}%
\providecommand \@ifx [1]{%
 \ifx #1\expandafter \@firstoftwo
 \else \expandafter \@secondoftwo
 \fi
}%
\providecommand \natexlab [1]{#1}%
\providecommand \enquote  [1]{``#1''}%
\providecommand \bibnamefont  [1]{#1}%
\providecommand \bibfnamefont [1]{#1}%
\providecommand \citenamefont [1]{#1}%
\providecommand \href@noop [0]{\@secondoftwo}%
\providecommand \href [0]{\begingroup \@sanitize@url \@href}%
\providecommand \@href[1]{\@@startlink{#1}\@@href}%
\providecommand \@@href[1]{\endgroup#1\@@endlink}%
\providecommand \@sanitize@url [0]{\catcode `\\12\catcode `\$12\catcode
  `\&12\catcode `\#12\catcode `\^12\catcode `\_12\catcode `\%12\relax}%
\providecommand \@@startlink[1]{}%
\providecommand \@@endlink[0]{}%
\providecommand \url  [0]{\begingroup\@sanitize@url \@url }%
\providecommand \@url [1]{\endgroup\@href {#1}{\urlprefix }}%
\providecommand \urlprefix  [0]{URL }%
\providecommand \Eprint [0]{\href }%
\providecommand \doibase [0]{http://dx.doi.org/}%
\providecommand \selectlanguage [0]{\@gobble}%
\providecommand \bibinfo  [0]{\@secondoftwo}%
\providecommand \bibfield  [0]{\@secondoftwo}%
\providecommand \translation [1]{[#1]}%
\providecommand \BibitemOpen [0]{}%
\providecommand \bibitemStop [0]{}%
\providecommand \bibitemNoStop [0]{.\EOS\space}%
\providecommand \EOS [0]{\spacefactor3000\relax}%
\providecommand \BibitemShut  [1]{\csname bibitem#1\endcsname}%
\let\auto@bib@innerbib\@empty
\bibitem [{\citenamefont {Franke-Arnold}\ \emph {et~al.}(2008)\citenamefont
  {Franke-Arnold}, \citenamefont {Allen},\ and\ \citenamefont
  {Padgett}}]{franke2008advances}%
  \BibitemOpen
  \bibfield  {author} {\bibinfo {author} {\bibfnamefont {S.}~\bibnamefont
  {Franke-Arnold}}, \bibinfo {author} {\bibfnamefont {L.}~\bibnamefont
  {Allen}}, \ and\ \bibinfo {author} {\bibfnamefont {M.}~\bibnamefont
  {Padgett}},\ }\href@noop {} {\bibfield  {journal} {\bibinfo  {journal} {Laser
  \& Photonics Reviews}\ }\textbf {\bibinfo {volume} {2}},\ \bibinfo {pages}
  {299} (\bibinfo {year} {2008})}\BibitemShut {NoStop}%
\bibitem [{\citenamefont {Bliokh}\ and\ \citenamefont
  {Nori}(2015)}]{BLIOKH2015physrep}%
  \BibitemOpen
  \bibfield  {author} {\bibinfo {author} {\bibfnamefont {K.~Y.}\ \bibnamefont
  {Bliokh}}\ and\ \bibinfo {author} {\bibfnamefont {F.}~\bibnamefont {Nori}},\
  }\href {\doibase https://doi.org/10.1016/j.physrep.2015.06.003} {\bibfield
  {journal} {\bibinfo  {journal} {Phys. Rep.}\ }\textbf {\bibinfo {volume}
  {592}},\ \bibinfo {pages} {1 } (\bibinfo {year} {2015})}\BibitemShut
  {NoStop}%
\bibitem [{\citenamefont {Bliokh}\ \emph
  {et~al.}(2015{\natexlab{a}})\citenamefont {Bliokh}, \citenamefont
  {Rodriguez-Fortuno}, \citenamefont {Nori},\ and\ \citenamefont
  {Zayats}}]{Bliokh2015Natpho}%
  \BibitemOpen
  \bibfield  {author} {\bibinfo {author} {\bibfnamefont {K.~Y.}\ \bibnamefont
  {Bliokh}}, \bibinfo {author} {\bibfnamefont {F.~J.}\ \bibnamefont
  {Rodriguez-Fortuno}}, \bibinfo {author} {\bibfnamefont {F.}~\bibnamefont
  {Nori}}, \ and\ \bibinfo {author} {\bibfnamefont {A.~V.}\ \bibnamefont
  {Zayats}},\ }\href {\doibase 10.1038/NPHOTON.2010.201} {\bibfield  {journal}
  {\bibinfo  {journal} {Nat. Photonics}\ }\textbf {\bibinfo {volume} {9}},\
  \bibinfo {pages} {796} (\bibinfo {year} {2015}{\natexlab{a}})}\BibitemShut
  {NoStop}%
\bibitem [{\citenamefont {Onoda}\ \emph {et~al.}(2004)\citenamefont {Onoda},
  \citenamefont {Murakami},\ and\ \citenamefont {Nagaosa}}]{Onoda2004}%
  \BibitemOpen
  \bibfield  {author} {\bibinfo {author} {\bibfnamefont {M.}~\bibnamefont
  {Onoda}}, \bibinfo {author} {\bibfnamefont {S.}~\bibnamefont {Murakami}}, \
  and\ \bibinfo {author} {\bibfnamefont {N.}~\bibnamefont {Nagaosa}},\ }\href
  {\doibase 10.1103/PhysRevLett.93.083901} {\bibfield  {journal} {\bibinfo
  {journal} {Phys. Rev. Lett.}\ }\textbf {\bibinfo {volume} {93}},\ \bibinfo
  {pages} {083901} (\bibinfo {year} {2004})}\BibitemShut {NoStop}%
\bibitem [{\citenamefont {Berry}\ \emph {et~al.}(2005)\citenamefont {Berry},
  \citenamefont {Jeffrey},\ and\ \citenamefont {Mansuripur}}]{Berry2005}%
  \BibitemOpen
  \bibfield  {author} {\bibinfo {author} {\bibfnamefont {M.~V.}\ \bibnamefont
  {Berry}}, \bibinfo {author} {\bibfnamefont {M.~R.}\ \bibnamefont {Jeffrey}},
  \ and\ \bibinfo {author} {\bibfnamefont {M.}~\bibnamefont {Mansuripur}},\
  }\href {http://stacks.iop.org/1464-4258/7/i=11/a=011} {\bibfield  {journal}
  {\bibinfo  {journal} {J. Opt. A: Pure Appl. Opt.}\ }\textbf {\bibinfo
  {volume} {7}},\ \bibinfo {pages} {685} (\bibinfo {year} {2005})}\BibitemShut
  {NoStop}%
\bibitem [{\citenamefont {Marrucci}\ \emph {et~al.}(2006)\citenamefont
  {Marrucci}, \citenamefont {Manzo},\ and\ \citenamefont
  {Paparo}}]{Marrucci2006}%
  \BibitemOpen
  \bibfield  {author} {\bibinfo {author} {\bibfnamefont {L.}~\bibnamefont
  {Marrucci}}, \bibinfo {author} {\bibfnamefont {C.}~\bibnamefont {Manzo}}, \
  and\ \bibinfo {author} {\bibfnamefont {D.}~\bibnamefont {Paparo}},\ }\href
  {\doibase 10.1103/PhysRevLett.96.163905} {\bibfield  {journal} {\bibinfo
  {journal} {Phys. Rev. Lett.}\ }\textbf {\bibinfo {volume} {96}},\ \bibinfo
  {pages} {163905} (\bibinfo {year} {2006})}\BibitemShut {NoStop}%
\bibitem [{\citenamefont {Zhao}\ \emph {et~al.}(2007)\citenamefont {Zhao},
  \citenamefont {Edgar}, \citenamefont {Jeffries}, \citenamefont {McGloin},\
  and\ \citenamefont {Chiu}}]{zhao2007}%
  \BibitemOpen
  \bibfield  {author} {\bibinfo {author} {\bibfnamefont {Y.}~\bibnamefont
  {Zhao}}, \bibinfo {author} {\bibfnamefont {J.~S.}\ \bibnamefont {Edgar}},
  \bibinfo {author} {\bibfnamefont {G.~D.~M.}\ \bibnamefont {Jeffries}},
  \bibinfo {author} {\bibfnamefont {D.}~\bibnamefont {McGloin}}, \ and\
  \bibinfo {author} {\bibfnamefont {D.~T.}\ \bibnamefont {Chiu}},\ }\href
  {\doibase 10.1103/PhysRevLett.99.073901} {\bibfield  {journal} {\bibinfo
  {journal} {Phys. Rev. Lett.}\ }\textbf {\bibinfo {volume} {99}},\ \bibinfo
  {pages} {073901} (\bibinfo {year} {2007})}\BibitemShut {NoStop}%
\bibitem [{\citenamefont {Hosten}\ and\ \citenamefont
  {Kwiat}(2008)}]{Hosten787}%
  \BibitemOpen
  \bibfield  {author} {\bibinfo {author} {\bibfnamefont {O.}~\bibnamefont
  {Hosten}}\ and\ \bibinfo {author} {\bibfnamefont {P.}~\bibnamefont {Kwiat}},\
  }\href {\doibase 10.1126/science.1152697} {\bibfield  {journal} {\bibinfo
  {journal} {Science}\ }\textbf {\bibinfo {volume} {319}},\ \bibinfo {pages}
  {787} (\bibinfo {year} {2008})}\BibitemShut {NoStop}%
\bibitem [{\citenamefont {Bliokh}\ \emph {et~al.}(2008)\citenamefont {Bliokh},
  \citenamefont {Niv}, \citenamefont {Kleiner},\ and\ \citenamefont
  {Hasman}}]{Bliokh2008}%
  \BibitemOpen
  \bibfield  {author} {\bibinfo {author} {\bibfnamefont {K.~Y.}\ \bibnamefont
  {Bliokh}}, \bibinfo {author} {\bibfnamefont {A.}~\bibnamefont {Niv}},
  \bibinfo {author} {\bibfnamefont {V.}~\bibnamefont {Kleiner}}, \ and\
  \bibinfo {author} {\bibfnamefont {E.}~\bibnamefont {Hasman}},\ }\href
  {http://dx.doi.org/10.1038/nphoton.2008.229} {\bibfield  {journal} {\bibinfo
  {journal} {Nat. Photonics}\ }\textbf {\bibinfo {volume} {2}},\ \bibinfo
  {pages} {748} (\bibinfo {year} {2008})}\BibitemShut {NoStop}%
\bibitem [{\citenamefont {Rodr\'{\i}guez-Herrera}\ \emph
  {et~al.}(2010)\citenamefont {Rodr\'{\i}guez-Herrera}, \citenamefont {Lara},
  \citenamefont {Bliokh}, \citenamefont {Ostrovskaya},\ and\ \citenamefont
  {Dainty}}]{rodriguez2010}%
  \BibitemOpen
  \bibfield  {author} {\bibinfo {author} {\bibfnamefont {O.~G.}\ \bibnamefont
  {Rodr\'{\i}guez-Herrera}}, \bibinfo {author} {\bibfnamefont {D.}~\bibnamefont
  {Lara}}, \bibinfo {author} {\bibfnamefont {K.~Y.}\ \bibnamefont {Bliokh}},
  \bibinfo {author} {\bibfnamefont {E.~A.}\ \bibnamefont {Ostrovskaya}}, \ and\
  \bibinfo {author} {\bibfnamefont {C.}~\bibnamefont {Dainty}},\ }\href
  {\doibase 10.1103/PhysRevLett.104.253601} {\bibfield  {journal} {\bibinfo
  {journal} {Phys. Rev. Lett.}\ }\textbf {\bibinfo {volume} {104}},\ \bibinfo
  {pages} {253601} (\bibinfo {year} {2010})}\BibitemShut {NoStop}%
\bibitem [{\citenamefont {Vuong}\ \emph {et~al.}(2010)\citenamefont {Vuong},
  \citenamefont {Adam}, \citenamefont {Brok}, \citenamefont {Planken},\ and\
  \citenamefont {Urbach}}]{Vuong2010}%
  \BibitemOpen
  \bibfield  {author} {\bibinfo {author} {\bibfnamefont {L.~T.}\ \bibnamefont
  {Vuong}}, \bibinfo {author} {\bibfnamefont {A.~J.~L.}\ \bibnamefont {Adam}},
  \bibinfo {author} {\bibfnamefont {J.~M.}\ \bibnamefont {Brok}}, \bibinfo
  {author} {\bibfnamefont {P.~C.~M.}\ \bibnamefont {Planken}}, \ and\ \bibinfo
  {author} {\bibfnamefont {H.~P.}\ \bibnamefont {Urbach}},\ }\href {\doibase
  10.1103/PhysRevLett.104.083903} {\bibfield  {journal} {\bibinfo  {journal}
  {Phys. Rev. Lett.}\ }\textbf {\bibinfo {volume} {104}},\ \bibinfo {pages}
  {083903} (\bibinfo {year} {2010})}\BibitemShut {NoStop}%
\bibitem [{\citenamefont {Lin}\ \emph {et~al.}(2013)\citenamefont {Lin},
  \citenamefont {Mueller}, \citenamefont {Wang}, \citenamefont {Yuan},
  \citenamefont {Antoniou}, \citenamefont {Yuan},\ and\ \citenamefont
  {Capasso}}]{Lin331}%
  \BibitemOpen
  \bibfield  {author} {\bibinfo {author} {\bibfnamefont {J.}~\bibnamefont
  {Lin}}, \bibinfo {author} {\bibfnamefont {J.~P.~B.}\ \bibnamefont {Mueller}},
  \bibinfo {author} {\bibfnamefont {Q.}~\bibnamefont {Wang}}, \bibinfo {author}
  {\bibfnamefont {G.}~\bibnamefont {Yuan}}, \bibinfo {author} {\bibfnamefont
  {N.}~\bibnamefont {Antoniou}}, \bibinfo {author} {\bibfnamefont {X.-C.}\
  \bibnamefont {Yuan}}, \ and\ \bibinfo {author} {\bibfnamefont
  {F.}~\bibnamefont {Capasso}},\ }\href {\doibase 10.1126/science.1233746}
  {\bibfield  {journal} {\bibinfo  {journal} {Science}\ }\textbf {\bibinfo
  {volume} {340}},\ \bibinfo {pages} {331} (\bibinfo {year}
  {2013})}\BibitemShut {NoStop}%
\bibitem [{\citenamefont {Shitrit}\ \emph {et~al.}(2013)\citenamefont
  {Shitrit}, \citenamefont {Yulevich}, \citenamefont {Maguid}, \citenamefont
  {Ozeri}, \citenamefont {Veksler}, \citenamefont {Kleiner},\ and\
  \citenamefont {Hasman}}]{Shitrit724}%
  \BibitemOpen
  \bibfield  {author} {\bibinfo {author} {\bibfnamefont {N.}~\bibnamefont
  {Shitrit}}, \bibinfo {author} {\bibfnamefont {I.}~\bibnamefont {Yulevich}},
  \bibinfo {author} {\bibfnamefont {E.}~\bibnamefont {Maguid}}, \bibinfo
  {author} {\bibfnamefont {D.}~\bibnamefont {Ozeri}}, \bibinfo {author}
  {\bibfnamefont {D.}~\bibnamefont {Veksler}}, \bibinfo {author} {\bibfnamefont
  {V.}~\bibnamefont {Kleiner}}, \ and\ \bibinfo {author} {\bibfnamefont
  {E.}~\bibnamefont {Hasman}},\ }\href {\doibase 10.1126/science.1234892}
  {\bibfield  {journal} {\bibinfo  {journal} {Science}\ }\textbf {\bibinfo
  {volume} {340}},\ \bibinfo {pages} {724} (\bibinfo {year}
  {2013})}\BibitemShut {NoStop}%
\bibitem [{\citenamefont {Petersen}\ \emph {et~al.}(2014)\citenamefont
  {Petersen}, \citenamefont {Volz},\ and\ \citenamefont
  {Rauschenbeutel}}]{Petersen67}%
  \BibitemOpen
  \bibfield  {author} {\bibinfo {author} {\bibfnamefont {J.}~\bibnamefont
  {Petersen}}, \bibinfo {author} {\bibfnamefont {J.}~\bibnamefont {Volz}}, \
  and\ \bibinfo {author} {\bibfnamefont {A.}~\bibnamefont {Rauschenbeutel}},\
  }\href {\doibase 10.1126/science.1257671} {\bibfield  {journal} {\bibinfo
  {journal} {Science}\ }\textbf {\bibinfo {volume} {346}},\ \bibinfo {pages}
  {67} (\bibinfo {year} {2014})}\BibitemShut {NoStop}%
\bibitem [{\citenamefont {O’Connor}\ \emph {et~al.}(2014)\citenamefont
  {O’Connor}, \citenamefont {Ginzburg}, \citenamefont {Rodríguez-Fortuño},
  \citenamefont {Wurtz},\ and\ \citenamefont {Zayats}}]{O’Connor2014}%
  \BibitemOpen
  \bibfield  {author} {\bibinfo {author} {\bibfnamefont {D.}~\bibnamefont
  {O’Connor}}, \bibinfo {author} {\bibfnamefont {P.}~\bibnamefont
  {Ginzburg}}, \bibinfo {author} {\bibfnamefont {F.~J.}\ \bibnamefont
  {Rodríguez-Fortuño}}, \bibinfo {author} {\bibfnamefont {G.~A.}\
  \bibnamefont {Wurtz}}, \ and\ \bibinfo {author} {\bibfnamefont {A.~V.}\
  \bibnamefont {Zayats}},\ }\href {http://dx.doi.org/10.1038/ncomms6327}
  {\bibfield  {journal} {\bibinfo  {journal} {Nat. Commun.}\ }\textbf {\bibinfo
  {volume} {5}},\ \bibinfo {pages} {5327} (\bibinfo {year} {2014})}\BibitemShut
  {NoStop}%
\bibitem [{\citenamefont {Bliokh}\ \emph
  {et~al.}(2015{\natexlab{b}})\citenamefont {Bliokh}, \citenamefont
  {Smirnova},\ and\ \citenamefont {Nori}}]{Bliokh1448}%
  \BibitemOpen
  \bibfield  {author} {\bibinfo {author} {\bibfnamefont {K.~Y.}\ \bibnamefont
  {Bliokh}}, \bibinfo {author} {\bibfnamefont {D.}~\bibnamefont {Smirnova}}, \
  and\ \bibinfo {author} {\bibfnamefont {F.}~\bibnamefont {Nori}},\ }\href
  {\doibase 10.1126/science.aaa9519} {\bibfield  {journal} {\bibinfo  {journal}
  {Science}\ }\textbf {\bibinfo {volume} {348}},\ \bibinfo {pages} {1448}
  (\bibinfo {year} {2015}{\natexlab{b}})}\BibitemShut {NoStop}%
\bibitem [{\citenamefont {Bliokh}(2006)}]{Bliokh2006PRL}%
  \BibitemOpen
  \bibfield  {author} {\bibinfo {author} {\bibfnamefont {K.~Y.}\ \bibnamefont
  {Bliokh}},\ }\href {\doibase 10.1103/PhysRevLett.97.043901} {\bibfield
  {journal} {\bibinfo  {journal} {Phys. Rev. Lett.}\ }\textbf {\bibinfo
  {volume} {97}},\ \bibinfo {pages} {043901} (\bibinfo {year}
  {2006})}\BibitemShut {NoStop}%
\bibitem [{\citenamefont {Ross}(1984)}]{Ross1984}%
  \BibitemOpen
  \bibfield  {author} {\bibinfo {author} {\bibfnamefont {J.~N.}\ \bibnamefont
  {Ross}},\ }\href {\doibase 10.1007/BF00619638} {\bibfield  {journal}
  {\bibinfo  {journal} {Opt. Quantum. Electron.}\ }\textbf {\bibinfo {volume}
  {16}},\ \bibinfo {pages} {455} (\bibinfo {year} {1984})}\BibitemShut
  {NoStop}%
\bibitem [{\citenamefont {Haldane}(1986)}]{Haldane:86}%
  \BibitemOpen
  \bibfield  {author} {\bibinfo {author} {\bibfnamefont {F.~D.~M.}\
  \bibnamefont {Haldane}},\ }\href {\doibase 10.1364/OL.11.000730} {\bibfield
  {journal} {\bibinfo  {journal} {Opt. Lett.}\ }\textbf {\bibinfo {volume}
  {11}},\ \bibinfo {pages} {730} (\bibinfo {year} {1986})}\BibitemShut
  {NoStop}%
\bibitem [{\citenamefont {Chiao}\ and\ \citenamefont {Wu}(1986)}]{Chiao1986}%
  \BibitemOpen
  \bibfield  {author} {\bibinfo {author} {\bibfnamefont {R.~Y.}\ \bibnamefont
  {Chiao}}\ and\ \bibinfo {author} {\bibfnamefont {Y.-S.}\ \bibnamefont {Wu}},\
  }\href {\doibase 10.1103/PhysRevLett.57.933} {\bibfield  {journal} {\bibinfo
  {journal} {Phys. Rev. Lett.}\ }\textbf {\bibinfo {volume} {57}},\ \bibinfo
  {pages} {933} (\bibinfo {year} {1986})}\BibitemShut {NoStop}%
\bibitem [{\citenamefont {Berry}(1987)}]{berry1987interpreting}%
  \BibitemOpen
  \bibfield  {author} {\bibinfo {author} {\bibfnamefont {M.}~\bibnamefont
  {Berry}},\ }\href@noop {} {\bibfield  {journal} {\bibinfo  {journal}
  {Nature}\ }\textbf {\bibinfo {volume} {326}},\ \bibinfo {pages} {277}
  (\bibinfo {year} {1987})}\BibitemShut {NoStop}%
\bibitem [{\citenamefont {Bliokh}(2009)}]{Bliokh2009JOA}%
  \BibitemOpen
  \bibfield  {author} {\bibinfo {author} {\bibfnamefont {K.~Y.}\ \bibnamefont
  {Bliokh}},\ }\href {http://stacks.iop.org/1464-4258/11/i=9/a=094009}
  {\bibfield  {journal} {\bibinfo  {journal} {J. Opt. A: Pure Appl. Opt.}\
  }\textbf {\bibinfo {volume} {11}},\ \bibinfo {pages} {094009} (\bibinfo
  {year} {2009})}\BibitemShut {NoStop}%
\bibitem [{\citenamefont {Bliokh}\ and\ \citenamefont
  {Desyatnikov}(2009)}]{Bliokh2009PRA}%
  \BibitemOpen
  \bibfield  {author} {\bibinfo {author} {\bibfnamefont {K.~Y.}\ \bibnamefont
  {Bliokh}}\ and\ \bibinfo {author} {\bibfnamefont {A.~S.}\ \bibnamefont
  {Desyatnikov}},\ }\href {\doibase 10.1103/PhysRevA.79.011807} {\bibfield
  {journal} {\bibinfo  {journal} {Phys. Rev. A}\ }\textbf {\bibinfo {volume}
  {79}},\ \bibinfo {pages} {011807} (\bibinfo {year} {2009})}\BibitemShut
  {NoStop}%
\bibitem [{\citenamefont {da~Costa}(1981)}]{Costa1981}%
  \BibitemOpen
  \bibfield  {author} {\bibinfo {author} {\bibfnamefont {R.~C.~T.}\
  \bibnamefont {da~Costa}},\ }\href {\doibase 10.1103/PhysRevA.23.1982}
  {\bibfield  {journal} {\bibinfo  {journal} {Phys. Rev. A}\ }\textbf {\bibinfo
  {volume} {23}},\ \bibinfo {pages} {1982} (\bibinfo {year}
  {1981})}\BibitemShut {NoStop}%
\bibitem [{\citenamefont {Burgess}\ and\ \citenamefont
  {Jensen}(1993)}]{burgess1993}%
  \BibitemOpen
  \bibfield  {author} {\bibinfo {author} {\bibfnamefont {M.}~\bibnamefont
  {Burgess}}\ and\ \bibinfo {author} {\bibfnamefont {B.}~\bibnamefont
  {Jensen}},\ }\href {\doibase 10.1103/PhysRevA.48.1861} {\bibfield  {journal}
  {\bibinfo  {journal} {Phys. Rev. A}\ }\textbf {\bibinfo {volume} {48}},\
  \bibinfo {pages} {1861} (\bibinfo {year} {1993})}\BibitemShut {NoStop}%
\bibitem [{\citenamefont {Ouyang}\ \emph {et~al.}(1999)\citenamefont {Ouyang},
  \citenamefont {Mohta},\ and\ \citenamefont {Jaffe}}]{Ouyang1999}%
  \BibitemOpen
  \bibfield  {author} {\bibinfo {author} {\bibfnamefont {P.}~\bibnamefont
  {Ouyang}}, \bibinfo {author} {\bibfnamefont {V.}~\bibnamefont {Mohta}}, \
  and\ \bibinfo {author} {\bibfnamefont {R.}~\bibnamefont {Jaffe}},\ }\href
  {\doibase http://dx.doi.org/10.1006/aphy.1999.5935} {\bibfield  {journal}
  {\bibinfo  {journal} {Ann. Phys.}\ }\textbf {\bibinfo {volume} {275}},\
  \bibinfo {pages} {297 } (\bibinfo {year} {1999})}\BibitemShut {NoStop}%
\bibitem [{\citenamefont {Longhi}(2007)}]{Longhi:07}%
  \BibitemOpen
  \bibfield  {author} {\bibinfo {author} {\bibfnamefont {S.}~\bibnamefont
  {Longhi}},\ }\href {\doibase 10.1364/OL.32.002647} {\bibfield  {journal}
  {\bibinfo  {journal} {Opt. Lett.}\ }\textbf {\bibinfo {volume} {32}},\
  \bibinfo {pages} {2647} (\bibinfo {year} {2007})}\BibitemShut {NoStop}%
\bibitem [{\citenamefont {Batz}\ and\ \citenamefont
  {Peschel}(2008)}]{batz2008}%
  \BibitemOpen
  \bibfield  {author} {\bibinfo {author} {\bibfnamefont {S.}~\bibnamefont
  {Batz}}\ and\ \bibinfo {author} {\bibfnamefont {U.}~\bibnamefont {Peschel}},\
  }\href {\doibase 10.1103/PhysRevA.78.043821} {\bibfield  {journal} {\bibinfo
  {journal} {Phys. Rev. A}\ }\textbf {\bibinfo {volume} {78}},\ \bibinfo
  {pages} {043821} (\bibinfo {year} {2008})}\BibitemShut {NoStop}%
\bibitem [{\citenamefont {Taira}(2011)}]{TAIRA2011}%
  \BibitemOpen
  \bibfield  {author} {\bibinfo {author} {\bibfnamefont {H.}~\bibnamefont
  {Taira}},\ }\href {http://stacks.iop.org/0953-4075/44/i=19/a=195401}
  {\bibfield  {journal} {\bibinfo  {journal} {J. Phys. B: At., Mol. Opt.
  Phys.}\ }\textbf {\bibinfo {volume} {44}},\ \bibinfo {pages} {195401}
  (\bibinfo {year} {2011})}\BibitemShut {NoStop}%
\bibitem [{\citenamefont {Wang}\ \emph {et~al.}(2014)\citenamefont {Wang},
  \citenamefont {Du}, \citenamefont {Xu}, \citenamefont {Liu},\ and\
  \citenamefont {Zong}}]{yong-longwang2014}%
  \BibitemOpen
  \bibfield  {author} {\bibinfo {author} {\bibfnamefont {Y.-L.}\ \bibnamefont
  {Wang}}, \bibinfo {author} {\bibfnamefont {L.}~\bibnamefont {Du}}, \bibinfo
  {author} {\bibfnamefont {C.-T.}\ \bibnamefont {Xu}}, \bibinfo {author}
  {\bibfnamefont {X.-J.}\ \bibnamefont {Liu}}, \ and\ \bibinfo {author}
  {\bibfnamefont {H.-S.}\ \bibnamefont {Zong}},\ }\href {\doibase
  10.1103/PhysRevA.90.042117} {\bibfield  {journal} {\bibinfo  {journal} {Phys.
  Rev. A}\ }\textbf {\bibinfo {volume} {90}},\ \bibinfo {pages} {042117}
  (\bibinfo {year} {2014})}\BibitemShut {NoStop}%
\bibitem [{\citenamefont {Wang}\ \emph
  {et~al.}(2017{\natexlab{a}})\citenamefont {Wang}, \citenamefont {Jiang},\
  and\ \citenamefont {Zong}}]{yong-longWang2017}%
  \BibitemOpen
  \bibfield  {author} {\bibinfo {author} {\bibfnamefont {Y.-L.}\ \bibnamefont
  {Wang}}, \bibinfo {author} {\bibfnamefont {H.}~\bibnamefont {Jiang}}, \ and\
  \bibinfo {author} {\bibfnamefont {H.-S.}\ \bibnamefont {Zong}},\ }\href
  {\doibase 10.1103/PhysRevA.96.022116} {\bibfield  {journal} {\bibinfo
  {journal} {Phys. Rev. A}\ }\textbf {\bibinfo {volume} {96}},\ \bibinfo
  {pages} {022116} (\bibinfo {year} {2017}{\natexlab{a}})}\BibitemShut
  {NoStop}%
\bibitem [{\citenamefont {Schultheiss}\ \emph {et~al.}(2010)\citenamefont
  {Schultheiss}, \citenamefont {Batz}, \citenamefont {Szameit}, \citenamefont
  {Dreisow}, \citenamefont {Nolte}, \citenamefont {T\"unnermann}, \citenamefont
  {Longhi},\ and\ \citenamefont {Peschel}}]{Schultheiss2010}%
  \BibitemOpen
  \bibfield  {author} {\bibinfo {author} {\bibfnamefont {V.~H.}\ \bibnamefont
  {Schultheiss}}, \bibinfo {author} {\bibfnamefont {S.}~\bibnamefont {Batz}},
  \bibinfo {author} {\bibfnamefont {A.}~\bibnamefont {Szameit}}, \bibinfo
  {author} {\bibfnamefont {F.}~\bibnamefont {Dreisow}}, \bibinfo {author}
  {\bibfnamefont {S.}~\bibnamefont {Nolte}}, \bibinfo {author} {\bibfnamefont
  {A.}~\bibnamefont {T\"unnermann}}, \bibinfo {author} {\bibfnamefont
  {S.}~\bibnamefont {Longhi}}, \ and\ \bibinfo {author} {\bibfnamefont
  {U.}~\bibnamefont {Peschel}},\ }\href {\doibase
  10.1103/PhysRevLett.105.143901} {\bibfield  {journal} {\bibinfo  {journal}
  {Phys. Rev. Lett.}\ }\textbf {\bibinfo {volume} {105}},\ \bibinfo {pages}
  {143901} (\bibinfo {year} {2010})}\BibitemShut {NoStop}%
\bibitem [{\citenamefont {Bekenstein}\ \emph {et~al.}(2017)\citenamefont
  {Bekenstein}, \citenamefont {Kabessa}, \citenamefont {Sharabi}, \citenamefont
  {Tal}, \citenamefont {Engheta}, \citenamefont {Eisenstein}, \citenamefont
  {Agranat},\ and\ \citenamefont {Segev}}]{Bekenstein2017}%
  \BibitemOpen
  \bibfield  {author} {\bibinfo {author} {\bibfnamefont {R.}~\bibnamefont
  {Bekenstein}}, \bibinfo {author} {\bibfnamefont {Y.}~\bibnamefont {Kabessa}},
  \bibinfo {author} {\bibfnamefont {Y.}~\bibnamefont {Sharabi}}, \bibinfo
  {author} {\bibfnamefont {O.}~\bibnamefont {Tal}}, \bibinfo {author}
  {\bibfnamefont {N.}~\bibnamefont {Engheta}}, \bibinfo {author} {\bibfnamefont
  {G.}~\bibnamefont {Eisenstein}}, \bibinfo {author} {\bibfnamefont {A.~J.}\
  \bibnamefont {Agranat}}, \ and\ \bibinfo {author} {\bibfnamefont
  {M.}~\bibnamefont {Segev}},\ }\href
  {https://doi.org/10.1038/s41566-017-0008-0} {\bibfield  {journal} {\bibinfo
  {journal} {Nat. Photonics}\ }\textbf {\bibinfo {volume} {11}},\ \bibinfo
  {pages} {664} (\bibinfo {year} {2017})}\BibitemShut {NoStop}%
\bibitem [{\citenamefont {Landau}(2013)}]{landau2013classical}%
  \BibitemOpen
  \bibfield  {author} {\bibinfo {author} {\bibfnamefont {L.~D.}\ \bibnamefont
  {Landau}},\ }\href@noop {} {\emph {\bibinfo {title} {The classical theory of
  fields}}},\ Vol.~\bibinfo {volume} {2}\ (\bibinfo  {publisher} {Elsevier},\
  \bibinfo {year} {2013})\BibitemShut {NoStop}%
\bibitem [{\citenamefont {Nakahara}(2003)}]{nakahara2003geometry}%
  \BibitemOpen
  \bibfield  {author} {\bibinfo {author} {\bibfnamefont {M.}~\bibnamefont
  {Nakahara}},\ }\href@noop {} {\emph {\bibinfo {title} {Geometry, topology and
  physics}}}\ (\bibinfo  {publisher} {CRC Press},\ \bibinfo {year}
  {2003})\BibitemShut {NoStop}%
\bibitem [{\citenamefont {Schuster}\ and\ \citenamefont
  {Jaffe}(2003)}]{SCHUSTER2003132}%
  \BibitemOpen
  \bibfield  {author} {\bibinfo {author} {\bibfnamefont {P.}~\bibnamefont
  {Schuster}}\ and\ \bibinfo {author} {\bibfnamefont {R.}~\bibnamefont
  {Jaffe}},\ }\href {\doibase http://dx.doi.org/10.1016/S0003-4916(03)00080-0}
  {\bibfield  {journal} {\bibinfo  {journal} {Ann. Phys.}\ }\textbf {\bibinfo
  {volume} {307}},\ \bibinfo {pages} {132 } (\bibinfo {year}
  {2003})}\BibitemShut {NoStop}%
\bibitem [{\citenamefont {Wang}\ \emph
  {et~al.}(2017{\natexlab{b}})\citenamefont {Wang}, \citenamefont {Lai},
  \citenamefont {Wang}, \citenamefont {Zong},\ and\ \citenamefont
  {Chen}}]{wang2017quantum}%
  \BibitemOpen
  \bibfield  {author} {\bibinfo {author} {\bibfnamefont {Y.-L.}\ \bibnamefont
  {Wang}}, \bibinfo {author} {\bibfnamefont {M.-Y.}\ \bibnamefont {Lai}},
  \bibinfo {author} {\bibfnamefont {F.}~\bibnamefont {Wang}}, \bibinfo {author}
  {\bibfnamefont {H.-S.}\ \bibnamefont {Zong}}, \ and\ \bibinfo {author}
  {\bibfnamefont {Y.-F.}\ \bibnamefont {Chen}},\ }\href@noop {} {\bibfield
  {journal} {\bibinfo  {journal} {arXiv preprint arXiv:1711.05453}\ } (\bibinfo
  {year} {2017}{\natexlab{b}})}\BibitemShut {NoStop}%
\bibitem [{\citenamefont {Białynicki-Birula}(1994)}]{Białynicki1994}%
  \BibitemOpen
  \bibfield  {author} {\bibinfo {author} {\bibfnamefont {I.}~\bibnamefont
  {Białynicki-Birula}},\ }\href@noop {} {\bibfield  {journal} {\bibinfo
  {journal} {Acta Phys. Pol. A}\ }\textbf {\bibinfo {volume} {86}},\ \bibinfo
  {pages} {97} (\bibinfo {year} {1994})}\BibitemShut {NoStop}%
\bibitem [{\citenamefont {Chen}\ \emph {et~al.}(2008)\citenamefont {Chen},
  \citenamefont {L\"u}, \citenamefont {Sun}, \citenamefont {Wang},\ and\
  \citenamefont {Goldman}}]{chen2008}%
  \BibitemOpen
  \bibfield  {author} {\bibinfo {author} {\bibfnamefont {X.-S.}\ \bibnamefont
  {Chen}}, \bibinfo {author} {\bibfnamefont {X.-F.}\ \bibnamefont {L\"u}},
  \bibinfo {author} {\bibfnamefont {W.-M.}\ \bibnamefont {Sun}}, \bibinfo
  {author} {\bibfnamefont {F.}~\bibnamefont {Wang}}, \ and\ \bibinfo {author}
  {\bibfnamefont {T.}~\bibnamefont {Goldman}},\ }\href {\doibase
  10.1103/PhysRevLett.100.232002} {\bibfield  {journal} {\bibinfo  {journal}
  {Phys. Rev. Lett.}\ }\textbf {\bibinfo {volume} {100}},\ \bibinfo {pages}
  {232002} (\bibinfo {year} {2008})}\BibitemShut {NoStop}%
\bibitem [{\citenamefont {Wang}\ \emph {et~al.}(2015)\citenamefont {Wang},
  \citenamefont {Wong}, \citenamefont {Chen}, \citenamefont {Sun},\ and\
  \citenamefont {Zhang}}]{Wang2015}%
  \BibitemOpen
  \bibfield  {author} {\bibinfo {author} {\bibfnamefont {F.}~\bibnamefont
  {Wang}}, \bibinfo {author} {\bibfnamefont {C.~W.}\ \bibnamefont {Wong}},
  \bibinfo {author} {\bibfnamefont {X.~S.}\ \bibnamefont {Chen}}, \bibinfo
  {author} {\bibfnamefont {W.~M.}\ \bibnamefont {Sun}}, \ and\ \bibinfo
  {author} {\bibfnamefont {P.~M.}\ \bibnamefont {Zhang}},\ }\href {\doibase
  10.1007/s00601-015-0969-9} {\bibfield  {journal} {\bibinfo  {journal}
  {Few-Body Systems}\ }\textbf {\bibinfo {volume} {56}},\ \bibinfo {pages}
  {249} (\bibinfo {year} {2015})}\BibitemShut {NoStop}%
\end{thebibliography}%

\end{document}